\documentclass[11pt,twoside]{article}

\usepackage{asp2006}
\usepackage{epsf}
\usepackage{psfig}
\usepackage{lscape}
\usepackage{graphicx}
\def\degr{\hbox{$^\circ$}}

\begin{document}
\title{The Need For High Resolution In Studies Of The 3-D Magnetic Field Structure Of AGN Jets}   
\author{Shane P. O'Sullivan and Denise C. Gabuzda}   
\affil{Department of Physics, University College Cork, Ireland}    

\begin{abstract} 

We are using ``broadband'' (4.6 to 43 GHz) multi-frequency VLBA polarization observations of compact AGN to investigate the 3-D structure of their jet magnetic ({\bf B}) fields. Observing at several frequencies, separated by short and long intervals, enables reliable determination of the distribution of Faraday Rotation, and thereby the intrinsic {\bf B} field structure. Transverse Rotation Measure (RM) gradients were detected in the jets of 0954+658 and 1418+546, providing evidence for the presence of a helical {\bf B} field surrounding the jet.
The RM in the core regions of 2200+420 (BL Lac), 0954+658 and 1418+546 display different signs in different frequency-intervals (on different spatial scales); we suggest an explanation for this in terms of modest bends in a helical {\bf B} field surrounding their jets. In future, polarization observations with a combination of VSOP-2 at 8, 22 and 43 GHz and ground arrays at frequencies with corresponding resolution will help map out the distributions of Faraday rotation, spectral index and the 3-D {\bf B} field structure both across the jet and closer to the central engine, providing strong constraints for any jet {\bf B} field models.

\end{abstract}


\section{Introduction}   

AGN jets emit synchrotron radiation that is often highly linearly polarized, which
provides important information on the degree of order and
orientation of the magnetic ({\bf B}) field in these jets.
``Blazars'' have jets pointed close to our line of sight (LoS) and often
exhibit strong variability in total flux and linear polarization over a
broad range of frequencies from $\gamma$-ray to radio.

Recent MHD simulations have provided an almost complete explanation of how these jets are launched, accelerated and collimated close to the black hole, see \cite{MeierJapan} and references therein. The
dominant {\bf B} field structure on these scales is helical. It is possible that this
structure changes when the flow gets disrupted by shocks after a few
hundred Schwarzschild radii, but observational evidence of helical
fields on scales larger than this \citep{Asada2002, GabuzdaMurray2004, Mahmud2008} suggests that remnants of the
earlier {\bf B} field structure remain or that a current driven helical kink
instability is generated \citep{NakamuraJapan, Carey2008}.

Faraday Rotation manifests itself as a linear dependence of the observed polarization angle
($\chi_{obs}$) on the wavelength ($\lambda$) squared described by the formula $\chi_{obs}=\chi_0+RM\lambda^2$,
where the Rotation Measure (RM) is proportional to the integral of
the electron density and the dot product of the {\bf B} field along
the LoS with the path length $\vec{dl}$. Hence, a positive/negative RM
tells you that the LoS {\bf B} field is pointing
towards/away from the observer and combined with the corrected
polarization orientation we are provided with a 3-D view of the
{\bf B} field structure. However, if the rotating medium is mixed in with the emitting plasma then
internal Faraday rotation occurs and the above formula does not generally apply.
Theoretical models \citep{Burn1966, Cioffi1980} of internal FR in a spherical or cylindrical
region have shown that if a rotation of greater than $45\degr$ is
observed then the Faraday rotation must be external.

\begin{figure}
 \begin{minipage}[t]{5.0cm}
 \begin{center}
 \includegraphics[width=5.0cm,clip]{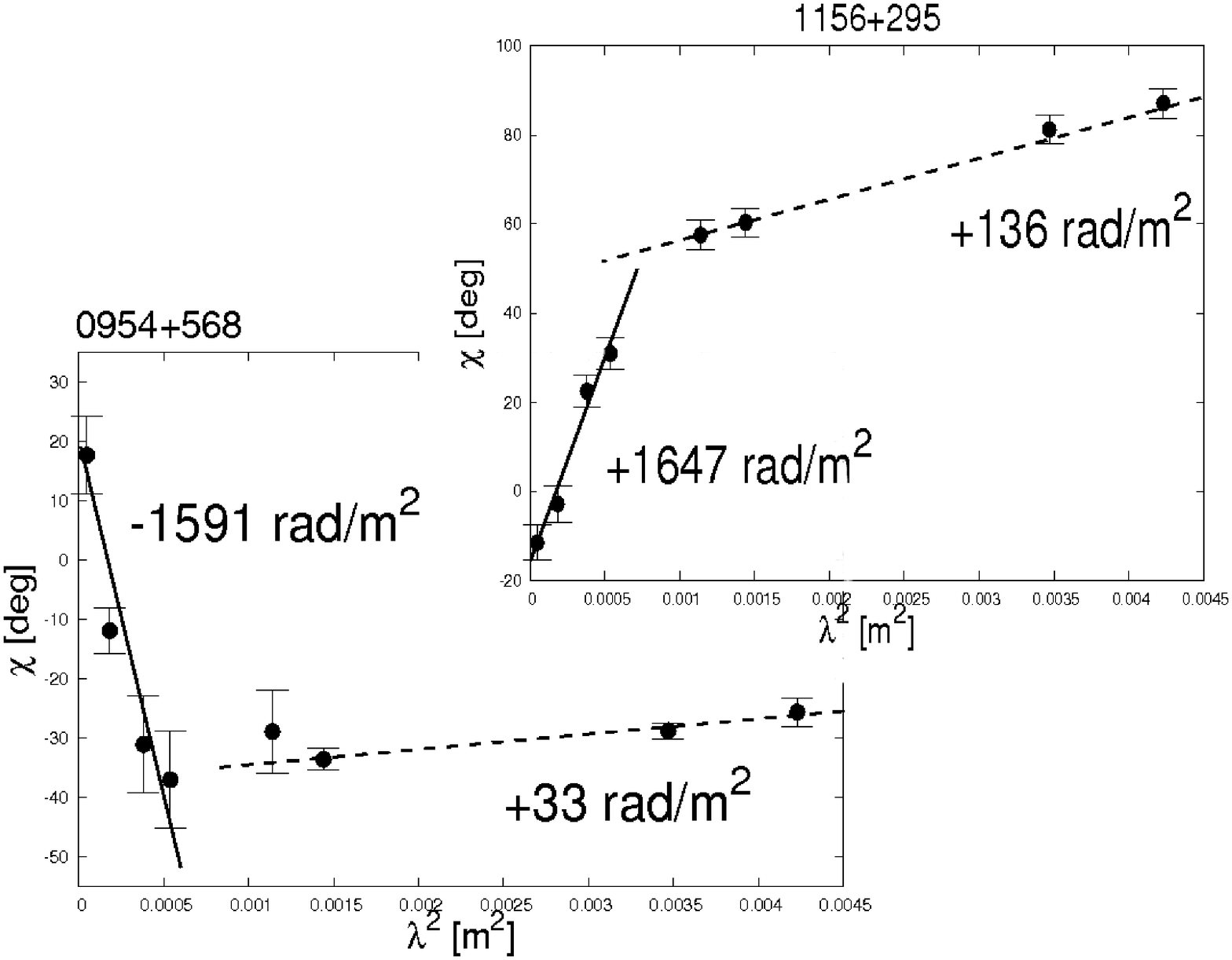}
 \end{center}
 \end{minipage}
 \hfill
 \begin{minipage}[t]{7.0cm}
 \begin{center}
 \includegraphics[width=6.5cm,clip]{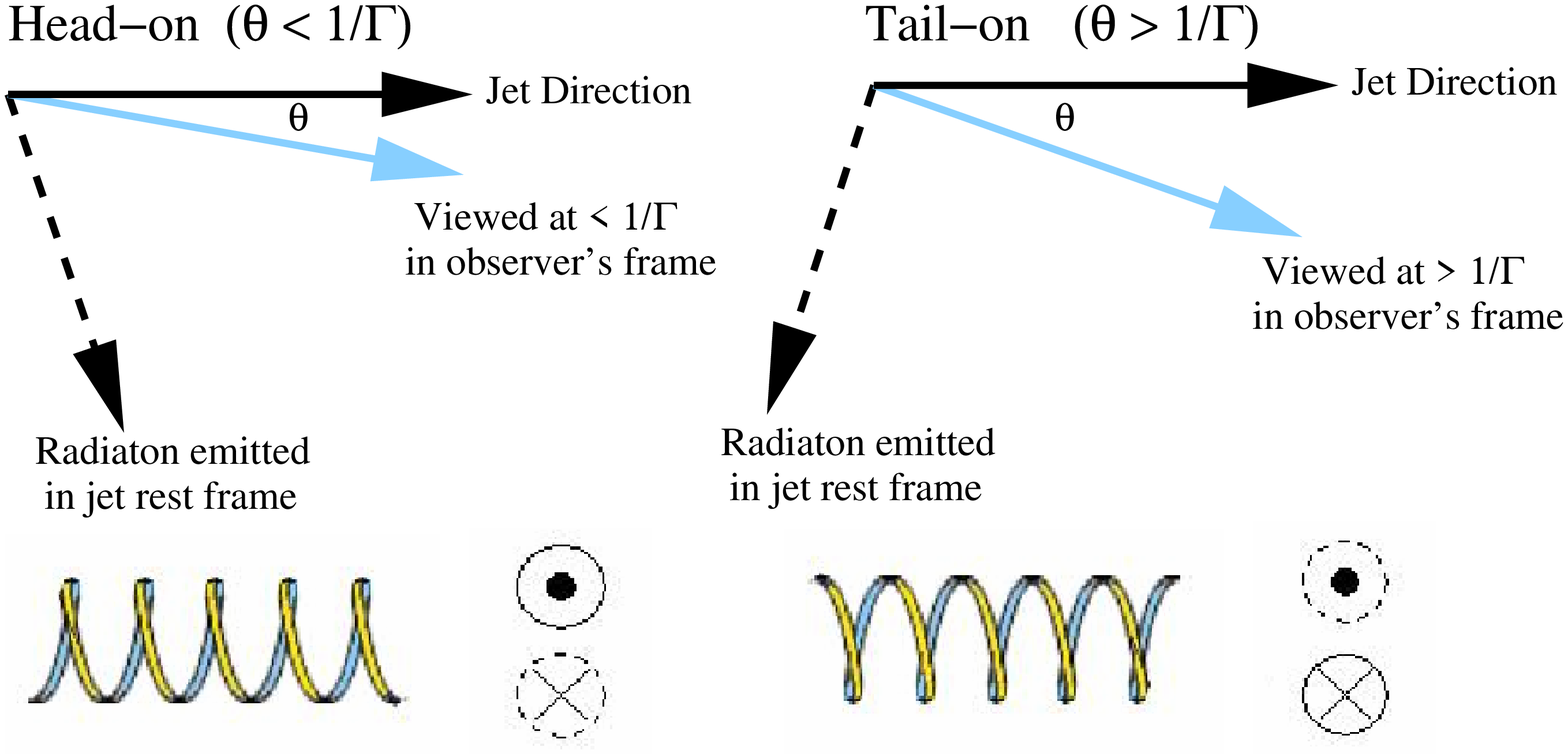}
 \end{center}
 \end{minipage}
 \caption{Left: Plots of $\chi$ vs. $\lambda^2$ for the whole observed frequency range of the core regions of 1156+295 and 0954+658 clearly showing a transition in both cases where physically different regions are being probed. Right: For a ``head-on'' ($\theta<1/\Gamma$) or ``tail-on'' ($\theta>1/\Gamma$) view of a helical {\bf B} field, different LoS components will dominate (shown by solid circles surrounding the dot or cross which indicate the direction of the LoS component) producing RMs with different signs in an unresolved jet.}
 \end{figure}

\section{Observations}
We observed 6 sources (Table 1) with the Very Long Baseline Array (VLBA) at 8 frequencies (4.6, 5.1, 7.9, 8.8, 12.9, 15.3, 22 \& 43 GHZ) over a 24-hr period on 2 July 2006.
All calibration and imaging was performed with the software
package AIPS. The integrated foreground (mainly Galactic) RMs were subtracted
using the multi-frequency VLA observations of \citet{Pushk2001}. This isolates the
RM distribution in the immediate vicinity of the AGN and is very important
for source rest frame calculations since the RM scales with $(1+z)^2$.

These observations are ideally suited for Faraday rotation analysis,
with both long and short spaced frequency intervals, which significantly
reduces the $\pm n\pi$ ambiguity in the observed polarization angles. This enables
us to construct RM maps on a range of different scales and Faraday depths,
providing a wealth of information on the RM distribution and its sign in the core
region and out along the jet. Matched-resolution images were constructed for
different frequency-intervals where clear transitions in RM occurred, corresponding to sampling
of physically different regions (eg. Figure 1 Left).

\section{Results}
The observed RMs are generally consistent with the rotating medium being external. Linear fits of $\chi$ vs. $\lambda^2$ are found in most cases with rotations of greater than $45\degr$ observed in several cases. Furthermore, the distribution of the degree of polarization of those with rotations less than $45\degr$ do not fit the corresponding predicted Faraday depolarization.
The degree of polarization in the core has a curious u-shaped distribution where it initially decreases rapidly from its value at the highest frequency, faster than predicted by beam depolarization \citep{ZTFog2}, and then after reaching a minimum increases at the lower frequencies. The increase in the degree of polarization at lower frequencies has a possible explanation in that higher polarized regions become blended with the core, as described in \citet{GabuzdaCawthorne1996} and \citet{Gabuzda1999}.

In all cases, the magnitude of the RM increases at higher frequencies, consistent with an increase in the electron density and/or {\bf B} field strength closer to the central engine. This can be quantified using the relation $RM \propto \nu^{a}$ derived in \citet{Jorstad2007}; see Table 1. Three of the sources are consistent with an outflow in a spherical or conical wind with $a\sim2$, similar to the results of \citet{Jorstad2007}. However, 0954+658 and 1418+546 have significantly higher values of 3.5 and 4.6, indicating much faster electron density falloffs in the Faraday rotating medium with distance from the central engine.

RM sign changes in different frequency-intervals are observed in 0954+658, 1418+546 and 2200+420.
In the case of 2200+420, the dominant jet {\bf B} field is transverse
throughout the jet, even as it bends, implying the presence of a global transverse
{\bf B} field structure. Bends in a jet, surrounded by a helical {\bf B} field, on scales smaller than is probed by the lowest frequency resolution will produce different dominant LoS {\bf B} field components due to the relativistic motion of the jet towards us, see Figure 1 Right and \citet{O'SullivanHEPRO}. This effect provides a natural explanation for the observed RM sign changes.
If there is no clear evidence for bends in the inner jet, then the same effect could be caused by an accelerating/decelerating jet
surrounded by a helical {\bf B} field; although it remains possible that bends are present on
scales smaller than our resolution.

\begin{table}
\caption{Summary of RM results. Intrinsic RM values in $rad/m^2$.}
\smallskip
\begin{center}
{\small
\begin{tabular}{cccccc}
\tableline
\noalign{\smallskip}
Blazar & z & Jet EVPA vs.& Core RM & Core RM & $a$\\
 & &Jet Direction& (Low $\nu$ range)& (High $\nu$ range) & \\
\noalign{\smallskip}
\tableline
\noalign{\smallskip}
$0954+658$ & $0.368$ & $\parallel$ & $+62\pm26$ & $-2977\pm496$ & $4.56\pm0.35$\\
\noalign{\smallskip}
$1156+295$ & $0.729$ & $\parallel$ & $+508\pm15$ & $+4983\pm475$ & $1.59\pm0.05$\\
\noalign{\smallskip}
$1418+546$ & $0.152$ & $\perp$ & $+100\pm15$ & $-577\pm64$ & $3.51\pm0.79$\\
\noalign{\smallskip}
 $1749+096$ & $0.320$ & $ $ & $-$ & $-$ & $-$\\
\noalign{\smallskip}
$2007+777$ & $0.342$ & $\parallel$ & $+1149\pm70$ & $+3429\pm229$ & $2.02\pm0.16$\\
\noalign{\smallskip}
$2200+420$ & $0.069$ & $\parallel$ & $+274\pm103$ & $-1152\pm49$ & $2.14\pm0.42$\\
\noalign{\smallskip}
\tableline
\end{tabular}
}
\end{center}
\end{table}


Further analysis of the RM structure of 0954+658 and 1418+546 reveals the presence of gradients in RM across the
jet, which are strong signatures for the presence of a helical {\bf B} field
geometry. For a side-on view of a helical {\bf B} field one would expect a
systematic gradient across the jet with the LoS {\bf B} field changing sign.
However, as was shown in \citet{Asada2002}, a positive asymmetric gradient (as
detected in 0954+658 and 1418+546) can be explained by having the viewing angle in the jet
rest frame less than the pitch angle of the helix.

\subsection{Intrinsic Polarization Orientation}

After correction for Faraday rotation, the jet polarization orientation is aligned with the jet direction in all sources except for 1418+546 where it's transverse to the jet direction. The intrinsic core polarization orientations vary with frequency-interval because the different RMs imply different zero-wavelength angles. Using the highest frequency-interval RM correction or the 43 GHz core polarization angle, the orientation of the furthest upstream polarized component is roughly transverse to the inner jet direction. Since these regions are optically thick in all cases, the {\bf B} field in the core is also transverse to the jet direction. This could be due to the toroidal component of a relatively tightly wound helical {\bf B} field and/or a strongly shocked region of plasma.

\section{Conclusions}
Polarization observations with VSOP-2 have the potential to greatly improve our understanding of the compact inner regions of AGN jets. The excellent resolution provided by VSOP-2, combined with an array of ground telescopes, will enable multi-frequency Faraday rotation analysis much closer to the central engine. The higher resolution of VSOP-2 will also reduce the effect of beam depolarization so that new polarization could be detected in previously unpolarized observations of AGN cores.

Observations using the full power of VSOP-2 with dual polarization at 8, 22 and 43 GHz combined with matched-resolution ground array observations at multiple bands at 15, 22, 43 and 86 GHz will facilitate unparalleled analysis of jet physics in these compact regions. Information on the 3-D {\bf B} field structure, electron density, 2-D internal structure and spectral index much closer to where these jets are being launched, accelerated and collimated will test current theoretical models of jet production and propagation through space.


 \acknowledgements 
 Funding for this research was provided by the Irish Research Council for Science, Engineering and Technology.


\end{document}